\documentclass[10pt,nofootinbib,twocolumn,aps,prd,preprintnumbers]{revtex4-1}
\usepackage{amsmath,amssymb,graphicx,bm,psfrag,color,slashed,subcaption,array,subfiles,cancel,bbm}
\usepackage[colorlinks=true, allcolors=blue]{hyperref}
\usepackage[dvipsnames]{xcolor}
\usepackage[normalem]{ulem}
\begin{document}

\preprint{IPPP/24/55}
% \vspace{0.5cm}
\title{On the Validity of Bounds on Light Axions for $f\lesssim10^{13}$ GeV}

\author{Martin Bauer}
\affiliation{Institute for Particle Physics Phenomenology, Department of Physics\\
Durham University, Durham, DH1 3LE, United Kingdom}

\author{Sreemanti Chakraborti}

\affiliation{Institute for Particle Physics Phenomenology, Department of Physics\\
Durham University, Durham, DH1 3LE, United Kingdom}

% e-mail addresses: one for each author, in the same order as the authors
%\emailAdd{martin.m.bauer@durham.ac.uk}
%\emailAdd{sreemanti.chakraborti@durham.ac.uk}

\begin{abstract}  
Light bosonic dark matter fields that can be treated like a classical wave have non-linear field values close to massive bodies. Here we make the important observation that the quadratic interactions of axion dark matter lead to non-perturbative axion field values for values of the decay constant of $f\lesssim 10^{13}$~GeV and masses $m_a\lesssim 7\times10^{8}/f\ {\rm eV^2}$ and generalise this result for axion-like particles. We identify experimental observables impacted by this effect.
\end{abstract}
\maketitle

\section{Introduction}
Ultralight bosons with masses below the eV scale are cold dark matter candidates with a very high occupation number, and can be described by classical waves~\cite{Hu:2000ke,Hui:2016ltb,Hui:2021tkt}. A particularly well motivated class of candidates are axions or more generally pseudo-Nambu Goldstone bosons resulting from the spontaneous breaking of a continuous, global symmetry, also called axion-like particles (ALPs). They are low-energy signatures in extensions of the Standard Model (SM) models in which an accidental global symmetry of the Standard Model, a different internal global symmetry, or a larger spacetime symmetry is spontaneously broken. 
This type of dark matter cannot be thermally produced but can be produced via the misalignment mechanism~\cite{Turner:1983sj, Turner:1983he}. As a result, they oscillate with an amplitude determined by the local dark matter density and their mass
\begin{align}\label{eq:wave}
a(t, \vec x)=\frac{\sqrt{2\rho_\text{DM}}}{m_a}\cos\, (m_a (t+\vec \beta\cdot\vec x))\,,
\end{align}
where $|\beta |\propto 10^{-3}$ is the virial dark matter velocity. The phenomenology of such a dark matter candidate differs significantly from particle-like dark matter candidates that could be observed with direct or indirect detection techniques. Instead, interactions between ultralight dark matter and the SM can induce oscillating variations of fundamental constants~\cite{Arvanitaki:2014faa}. Importantly, the behaviour of the dark matter wave depends on the local matter potential sourced by massive bodies such as the earth~\cite{Hees:2018fpg}. For quadratic interactions in the dark matter field, this source term induces an effective mass term near the massive body. Depending on the sign of the quadratic interaction term this results in a screening effect rendering ground-based experiments less sensitive or gives rise to an non-perturbative field values~\cite{Hees:2018fpg, Banerjee:2022sqg}.

In this letter, we show that the sign of the quadratic interaction is fixed for the axion and in almost all cases for axion-like particles with more general interactions and the axion field value becomes non-perturbative, generalising a result first derived in \cite{Hook:2017psm} for the case of a QCD axion that does not contribute to dark matter. As a result, many experimental constraints on light axions cannot be considered reliable. We determine the corresponding parameter space in terms of the axion decay constant and the axion couplings to SM particles and identify the experimental search strategies impacted by this effect as well as those that can avoid it. A more detailed analysis for the general case of an axion-like particle, taking into account effects from running and matching as well as a comparison between experimental sensitivities to linear and quadratic ALP interactions is presented in a companion paper~\cite{longpaper}.

\section{Quadratic Axion Interactions}

We consider an extension of the SM with a single pseudoscalar spin-0 field $a(x)$ that only interacts with the $SU(3)_C$ field strength tensor such that the axion Lagrangian can be written as
\begin{align}\label{eq:lag}
\mathcal{L}=\frac12 \partial_\mu a\,\partial^\mu a -\frac{m_a^2}{2}a^2+c_{GG}\frac{\alpha_s}{4\pi}\frac{a}{f}G_{\mu\nu} \tilde G^{\mu\nu}\,.
\end{align}
Factoring out $\alpha_s$ is a convenient choice that renders the coefficient $c_{GG}$ scale independent~\cite{Bauer:2021wjo}. An alternative, commonly used formulation in which the coefficient $c_{GG}$ is absorbed in the axion decay constant is defined as $f=-2c_{GG} f_a$~\cite{GrillidiCortona:2015jxo}. Below the QCD scale the interaction in \eqref{eq:lag} induces a correction to the pion mass term that can be derived from the chiral Lagrangian~\cite{Bauer:2020jbp} as
\begin{align}
m_{\pi,\text{eff}}^2(a)=m_\pi^2\bigg(1+\delta_\pi\frac{a^2}{f^2}\bigg)\,,\quad
\end{align}
with 
\begin{align}
\delta_\pi&=-2 c_{GG}^2\frac{m_u m_d}{(m_u+m_d)^2}\,.
\label{Eq:shift_mpi2}
\end{align}
This correction is strictly negative for any choice of $c_{GG}$. 
The axion is a pseudoscalar, but at the low scale it has scalar quadratic interactions with nucleons, electrons and photons described by the dimension six operators
\begin{align}\label{eq:a2Lag}
   {\cal L}_{\rm eff}^{D= 6}
  % &=\bar N\left(C_{N}(\mu)\mathbbm{1}+C_{\delta}(\mu)\tau\right)  N \frac{a^2}{f^2} \notag \\
  % &+C_{E}(\mu)\frac{a^2}{f^2} \bar e e  +  C_{\gamma}(\mu) \frac{a^2}{4f^2} F_{\mu\nu}F^{\mu\nu}\,,\\
    &= -M_{N}\,\delta_N \bar N N \frac{a^2}{f^2} +\Delta M_N \delta_{\Delta M} \bar N \tau N \frac{a^2}{f^2} \notag \\
   &-m_e\alpha \delta_e \bar e e  \frac{a^2}{f^2}  + \alpha\delta_\alpha  F_{\mu\nu}F^{\mu\nu}\frac{a^2}{4f^2}\,,
 \end{align}
with $\tau=\text{diag} (1,-1)$\,. 
One can then define the analogous corrections to the nucleon mass $M_N$, the neutron-proton mass difference $\Delta M_N = m_N-m_P$, the fine-structure constant $\alpha$ and the electron mass $m_e$ as
\begin{align}
M_N(a)&=M_N\bigg(1+\delta_N\frac{a^2}{f^2}\bigg)\,,\\
\Delta M_N(a)&=\Delta M\bigg(1+\delta_{\Delta M}\frac{a^2}{f^2}\bigg)\,,\\
\alpha^\text{eff}(a)&=\alpha\bigg(1+\delta_\alpha\frac{a^2}{f^2}\bigg)\,,\\
m_e(a)&=m_e\bigg(1+\delta_e\frac{a^2}{f^2}\bigg)\,,
\end{align}
with the coefficients~\cite{Banerjee:2022sqg, Kim:2023pvt, longpaper}
\begin{align}
 \delta_N &=-4c_1 \frac{m_\pi^2}{M_N} \delta_\pi,\label{eq:deltaN}\\ 
 \delta_{\Delta M}&=\delta_\pi\,,\label{eq:deltaDM}\\
 \delta_\alpha&=\frac{1}{12\pi} \left(1-32c_1\frac{m_\pi^2}{M_N}\right)\delta_\pi\,,\label{eq:deltaalpha}\\
  \delta_e&=\frac{\alpha}{16\pi^2} \mathrm{ln} \frac{m_e^2}{m_\pi^2}\bigg(1-32c_1\frac{m_\pi^2}{M_N}\bigg)\, \delta_\pi\,, \label{eq:deltae} 
\end{align}
where $c_1\approx -1.26$ ${\rm GeV^{-1}}$~\cite{Alarcon:2012kn}. Note that all coefficients in \eqref{eq:deltaN}-\eqref{eq:deltae} are proportional to $\delta_\pi$ and therefore negative as well. \footnote{We note that \cite{Beadle:2023flm} find the opposite sign for $\delta_\alpha$, which would affect the sign of $\delta_e$ here as well, but does not change our conclusions.}

Damour and Donoghue~\cite{Damour:2010rp} have shown that the effects of scalar couplings to matter made from atoms of mass $m_A$ can be captured by the coupling function
\begin{align}
\alpha_A &= \frac{\partial \ln m_A(a^2/f^2)}{\partial (a^2/f^2)}=\sum_i Q_i\delta_i\,,
\end{align} 
with $\delta_i=\delta_\pi, \delta_{\Delta M},\delta_e, $ and $\delta_\alpha$ and the corresponding `dilatonic charges' can be written in terms of the atomic mass number $A$ and charge $Z$ as
\begin{align}\label{eq:dilatoncharges}
\!Q_{\hat m}&=\!\left[9.3\!-\!\frac{3.6}{A^{1/3}}\!-\!2\frac{(A\!-\!2Z)^2}{A^2}\!-\!0.014\frac{Z(Z\!-\!1)}{A^{4/3}}\right]\!\times \!10^{-2}\!\!,\notag\\
\!Q_{\Delta M}&=1.7\times 10^{-3}\frac{A-2Z}{A}\,,\notag\\
\!Q_{\alpha}&=\left[-1.4+8.2\frac{Z}{A}+7.7\,\frac{Z(Z-1)}{A^{4/3}}\right]\!\times \!10^{-4}\,,\notag\\
\!Q_{e}&=5.5\times 10^{-4}\frac{Z}{A}\,,
\end{align}
where $Q_{\hat m}$ is proportional the sum of quark masses corresponding to $\delta_\pi$ and we have used the approximation $m_A= A M_N$ for the mass of the nucleus. For any realistic configuration~\eqref{eq:dilatoncharges} the dilatonic charges are strictly positive, such that 
\begin{align}
\alpha_A < 0\,
\end{align}
for the parameters \eqref{Eq:shift_mpi2} and \eqref{eq:deltaN}-\eqref{eq:deltae}. 

\section{Axion dark matter near massive bodies}
In the case of the axion, any linear interaction at low energy is
spin-dependent or suppressed by CP violating parameters and we can set a potential source term $J_\text{source}=0$ for non-polarised sources. 
The equation of motion for an axion near a massive source such as earth can therefore be written as
\begin{align}\label{eq:EOM}
(\partial_t^2-\Delta+m_a^2)a
&=-\sin\left(\frac{a}{f}\right)\sum_i\,\frac{Q_i^\text{source}\delta_i}{f}\rho_\text{source}\notag(r)\notag\\
&\kern-1.2cm=-\frac{a}{f}\sum_i\,\frac{Q_i^\text{source}\delta_i}{f}\rho_\text{source}(r)+\mathcal{O}\left(\frac{a^3}{f^3}\right)
\end{align}
where the effective source mass density $\rho_\text{source}(r)$ depends on the distance $r$ and the leading term in the second line is a valid approximation for small field values. The source term up to quadratic axion interactions can then be absorbed in the effective mass term~\cite{Hees:2018fpg, Banerjee:2022sqg}
\begin{align}\label{eq:EOMfull}
\bar m_a^2(r)=m_a^2+\sum_i \frac{Q_i^\text{source} \delta_i}{f^2}\rho_\text{source}(r)\,,
\end{align}
such that one can write the equations of motion as
\begin{align}
(\partial_t^2-\Delta+\bar m_a^2(r))a
&=0\,.
\end{align}

The solution of the equation of motion depends sensitively on the boundary conditions~\cite{Banerjee:2022sqg}. Under the assumption that the axion field at infinity takes the oscillating galactic background field value the authors of \cite{Hees:2018fpg} derived the solution for a generic scalar close to a spherical, massive body with radius $R_\text{source}$. In the case of the axion this soluiton reads
\begin{align}\label{eq:fullsol}
a(t, r) =  \frac{\sqrt{2\rho_\text{DM}}}{m_a} \cos(m_a t) \bigg[1 -  Z_\delta J_{\pm} \big( \sqrt{3|Z_\delta|}\big) \frac{R_\text{source}}{r}\bigg] ,
\end{align}
where the function $J_\pm(x)=J_{\text{sgn}(Z_\delta)}(x)$ depends on the sign of 
\begin{align}\label{eq:DMwave}
Z_\delta=\frac{1}{4\pi f^2}\frac{M_\text{source}}{R_\text{source}} \sum_iQ_i^\text{source}\delta_i\,,
\end{align}  
and
\begin{align}
J_+(x) &=\frac{3}{x^3}(x-\tanh x)\,,\notag\\
J_-(x)&=\frac{3}{x^3}(\tan x-x)\,.
\end{align}
Since the charges $Q_i^\text{source}$ are always positive and the parameters $\delta_i$ are negative, it follows for the case of the axion $Z_\delta <0$ and the solution \eqref{eq:DMwave} involves $J_-(x)$, independent of the sign of the axion coupling to gluons \eqref{eq:lag}.
The function $J_-(x)$ diverges for values of $x\to \pi/2$, and the axion field value is non-pertubative for values of $|Z_\delta|\approx \pi/3$. We approximate the matter content of earth as 1/3 iron ($A=56, Z=26$) and 2/3 silicon dioxide ($A=60, Z=30$) so that the charges for earth read
\begin{align}
\left\{Q^\oplus_{\hat m},Q^\oplus_{\delta m},Q^\oplus_\alpha,Q^\oplus_e \right\}\approx \left\{83,0.039, 0.26,0.27\right\}\times 10^{-3}\,, 
\end{align}
and the fraction of the mass and radius of the earth can be written as $M_\oplus/R_\oplus\approx 10^{29}\,\text{GeV}^2$. As a result, there is a critical value for the axion-gluon coupling 
\begin{align}
\frac{c_{GG}}{f}&\gtrsim \left(\frac{6}{\pi^3}\frac{m_u m_d}{(m_u+m_d)^2}\frac{M_\oplus}{R_\oplus} |Q_{\hat{m}}|\right)^{-1/2}\\[2pt]
&\approx 5.25\times 10^{-14}\,\text{GeV}^{-1}\,,
%\frac{3}{4\pi}\frac{M_\oplus}{R_\oplus} |Q_{\hat{m}}|\right)^{-1/2}\approx \frac{1}{10^{14}}\,\text{GeV}^{-1}\,,
\end{align}
and the axion field becomes large $a/f\approx 1$ for larger values of the interaction strength $c_{GG}/f$.  Note that this is not the case for general scalar dark matter with quadratic interactions for which the sign of the parameters $\delta_i$ is arbitrary and the positive solution $J_{+}(x)$ results in a screening for strong interactions~\cite{Hees:2018fpg, Banerjee:2022sqg}.

The axion field value diverges for large $x$ if the potential is truncated keeping only the leading term in \eqref{eq:EOM}. Since the axion potential is periodic, higher order operators in the expansion will regulate this divergence resulting in a cutoff for the field value of $a\sim f$. This effect was already observed in \cite{Hook:2017psm} in the case of non-dark matter axions and later discussed in detail in the context of astrophysical observables in \cite{Balkin:2020dsr, DiLuzio:2021pxd, Balkin:2022qer, Balkin:2023xtr}. An important difference in the case of dark matter is the boundary condition at $r\to \infty$, which in the case of~\cite{Hook:2017psm} corresponds to a vanishing field value, whereas in the case of axion dark matter it should be finite and non-vanishing. As a consequence of this boundary condition the axion mass dependence of the solution \eqref{eq:fullsol} is captured by the free oscillating field \eqref{eq:wave}. However, in order for the axion field value to deviate from the vacuum solution the potential energy induced by the source needs to be sufficiently large to turn the axion mass tachyonic~\cite{Hook:2017psm}. As a result, we only expect the axion to acquire large field values if
\begin{align}
m_a^2f^2+\frac{\sigma}{M_N}\rho_N \,\delta_\pi <0\,,
\end{align}
where we only kept the dominant term in the sum of \eqref{eq:DMwave}, $\rho_N$ denotes the nucleon mass density and $\sigma=-4c_1 m_\pi^2\approx 45$ MeV. In the case of earth as a source one finds 
\begin{align}\label{eq:ma2f2}
    m_a^2f^2< 5\times 10^{17}\,c_{GG}^2\,\text{eV}^4\,.
\end{align}

\begin{figure}[!ht]
    \centering
    \includegraphics[width=.55\textwidth]{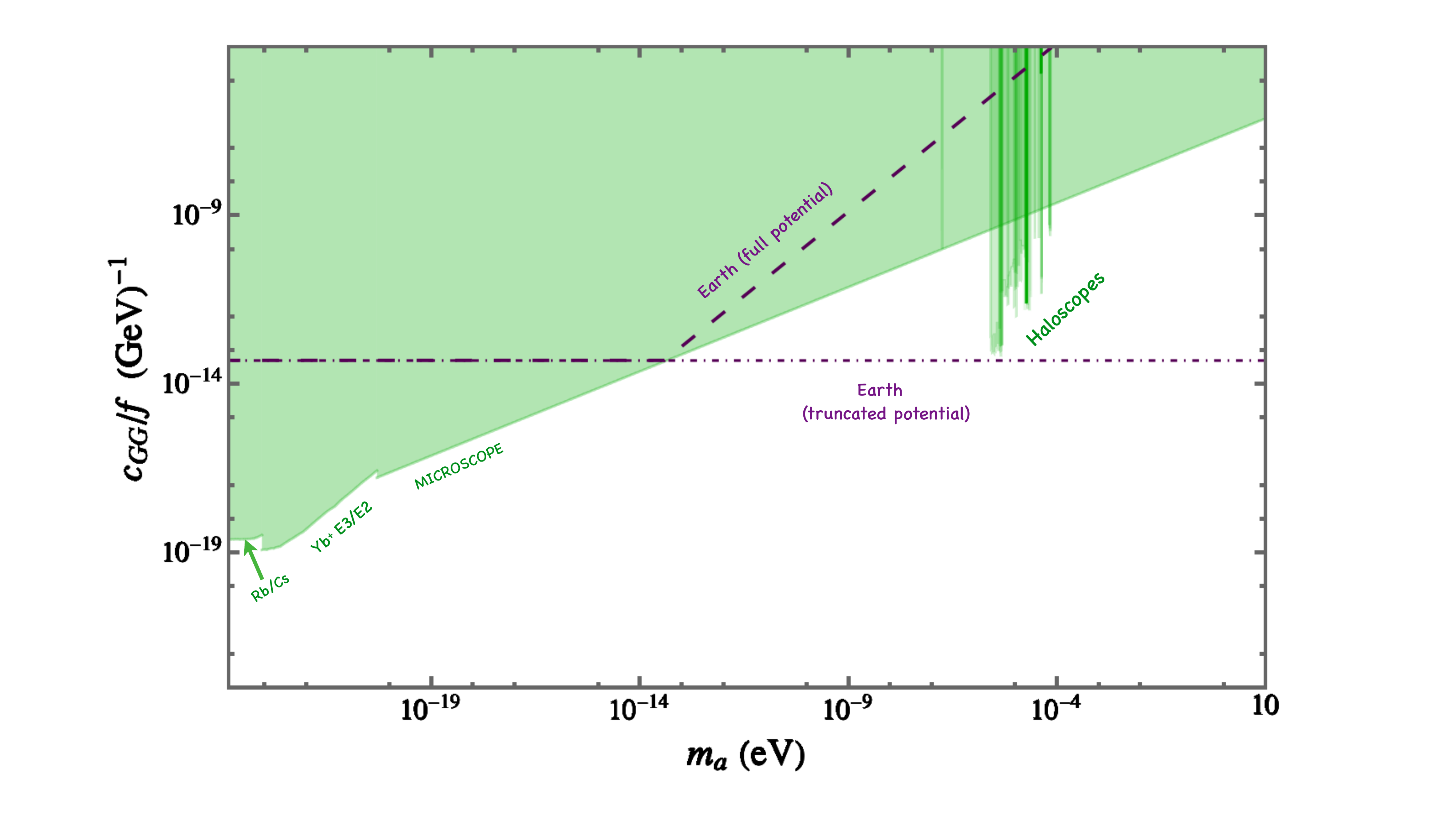}
    \caption{\small Bounds from Rb/Cs clocks, Yb$+$ clocks, MICROSCOPE and haloscope searches in green. The purple dashed line limits the non-perturbative parameter space expected from the solution using the full axion potential, the dot-dashed line represents the bound derived from truncating the potential at the quadratic order using dark matter boundary conditions. }
    \label{fig:comb}
\end{figure}

\section{Experiments}

Different experimental search strategies for axions are affected by the non-perturbative axion field values close to the surface of the earth for large values of $c_{GG}/f$. In particular results from experiments sensitive to the oscillations derived from \eqref{eq:wave} are not reliable in this parameter space.  This affects the results from nuclear magnetic resonance searches for spin precession~
\cite{Graham:2013gfa,Budker:2013hfa,JacksonKimball:2017elr,Agrawal:2022wjm}. 
Similarly, haloscopes that rely on the Primakoff effect to produce signal photons from the interaction between the local axion field and an external magnetic field are affected~\cite{Stern:2015kzo,ADMX:2019uok,McAllister:2023ipr}.  Similarly, light-shining-through-a-wall experiments rely on the local axion field value to produce a signal are impacted by the non-perturbative value close to the surface of the earth~\cite{Anselm:1985obz,VanBibber:1987rq}. 

Beyond searches for axion dark matter signals induced by linear axion interactions, also searches for quadratic axion couplings have to be reconsidered. This affects tests of the equivalence principle that rely on the gradient induced by the quadratic axion field~\cite{Touboul:2017grn, Touboul:2022yrw}. Probes on satellites in earth's orbit are less affected due to the suppression factor $R_\text{source}/r$ in \eqref{eq:DMwave}, but the parameter space for which the axion field has non-perturbative values just shifts towards larger axion couplings. Another class of probes that look for temporal variations of fundamental constants are sensitive to the local value of the axion field squared. This includes searches for oscillating modifications of transition frequencies in mechanical resonators~\cite{Branca:2016rez, Manley:2019vxy}, atomic and molecular clocks~\cite{Hees:2016gop, BACON:2020ubh, Flambaum:2023bnw, Filzinger:2023zrs, Kozlov:2013lha}, potential future nuclear clocks~\cite{Banerjee:2020kww, Banerjee:2023bjc,Caputo:2024doz}, laser interferometers~\cite{Grote:2019uvn,Savalle:2020vgz} and atom interferometers~\cite{Zhao:2021tie,Buchmueller:2023nll}. 

There are experimental searches for axions that do not rely on the local field value and are therefore not directly impacted by the non-perturbative field values discussed here. For example, helioscopes avoid this issue because they search for energetic axions produced in the sun~\cite{CAST:2017uph,Flambaum:2022zuq, McAllister:2022ibe}. The production of axions at colliders~\cite{Mimasu:2014nea,Bauer:2017ris,Brivio:2017ije} or in the decay of SM particles~\cite{Gavela:2019wzg, Bauer:2021mvw} can constrain the parameter space independent of the local axion field value. Similar conclusions can be drawn for cosmological phenomenon such as axinovae~\cite{Fox:2023xgx} which also remains unaffected.
In Fig.~\ref{fig:comb} we show a compilation of experimental constraints that would be impacted by non-perturbative field values of a dark matter axion field in green. The green shaded exclusions only provide reliable constraints for values of $c_{GG}/f\lesssim  5.25\times 10^{-14}\,\text{GeV}^{-1}$ and $m_a\lesssim 7\times10^{8}\,c_{GG}/f\ {\rm eV^2}$, outside the purple dashed line. We further show the region for which we expect the full solution to \eqref{eq:EOMfull} to develop a mass dependence according to \eqref{eq:ma2f2}. 

\section{Axionlike particles}

If additional couplings of the field $a(t,x)$ with the SM are present, the expressions for the parameters \eqref{Eq:shift_mpi2} and \eqref{eq:deltaN}-\eqref{eq:deltae} change. In particular, couplings to up and down quarks enter the expression for $\delta_\pi$, such that~\cite{longpaper} 
\begin{align}\label{eq:deltagen}
\delta_\pi&=-\frac{1}{8(1-\tau_a)^2}\bigg[4c_{GG}^2\Big(1-\tau_a^2-\frac{\Delta_m^2}{\hat m^2}(1-2\tau_a)\Big)\notag\\
&\qquad  +4c_{GG}(c_u-c_d)\frac{\Delta_m}{\hat m}\tau_a^2+(c_u-c_d)^2\tau_a^2\bigg]\,,
\end{align}
where $\tau_a=m_a^2/m_\pi^2$, $\hat m=m_u+m_d$ and $\Delta_m= m_d-m_u$ measures the isospin breaking. Note that all terms in \eqref{eq:deltagen}  proportional to the quark masses are suppressed by the mass of the axionlike particle. In the absence of the coupling to gluons the value of $\delta_\pi$ is still strictly negative, whereas the interference term can have either sign but is always subleading, such that couplings to quarks do not change our conclusions. One can ask whether there are potential interactions with SM particles that enter \eqref{eq:deltaN}-\eqref{eq:deltae} as contributions not proportional to $\delta_\pi$. In fact, a potential ALP-coupling to electrons can induce a contribution to $\delta_e$ which is independent of $\delta_\pi$. However, the shift-symmetry of the ALP leads to a heavily suppressed quadratic interaction.

\section{Conclusions}

We show that the structure of the axion Lagrangian leads to a universal negative sign for the quadratic axion interactions with matter. As a result, the field value for the axion close to a massive body like earth is non-perturbative for $f\lesssim  2\times 10^{13}\,\text{GeV}$ and $m_a\lesssim 7\times 10^{8}\,c_{GG}/f\ {\rm eV^2} $. This impacts a large range of experimental searches for axion dark matter that probes local gradients or temporal oscillations in electric dipole moments and fundamental constants.

The issue of non-perturbative field values for light scalars in the presence of massive bodies has first been observed in tensor–scalar theories of gravitation~\cite{Damour:1993hw, Damour:1996ke}. This phenomenon can be interpreted as \emph{spontaneous scalarization} in analogy to the spontaneous magnetization of ferromagnets, where the order parameter is the parameter $\alpha_A$ instead of the total magnetization. In the case of scalar extensions of general relativity the interaction strengths of the scalar is not a free parameter such that the non-perturbative regime depends solely on the mass of the central body~\cite{Damour:1996ke}. As a result, \emph{spontaneous scalarization} is expected for very massive bodies like neutron stars, whereas it is not an issue in the case of less massive sources like earth. For axions instead, non-perturbative solutions appear for any given set of charges as a function of its interaction strength. This has first been pointed out for the case of the QCD axion not contributing to dark matter by the authors of \cite{Hook:2017psm} and applied to derive constraints from neutron stars and other astrophysical observables in \cite{Balkin:2020dsr, DiLuzio:2021pxd, Balkin:2022qer, Balkin:2023xtr}. Here we calculate the axion couplings for which the non-perturbative field value develops in the case of axion dark matter and generalise it for the case of axion-like particles. A full solution of the Einstein and axion field equations for an oscillating dark matter field should reproduce the properties derived here. 

In the general case of light dark matter with quadratic couplings to SM matter the coefficients of the corresponding operators can have either sign and non-perturbative field values only occur for negative values of $\alpha_A$~\cite{Hees:2018fpg}. In the case of the axion quadratic interactions at low energy are proportional to the shift-symmetry breaking and therefore proportional to the pion mass shift. The sign of this mass shift is crucial for the non-perturbative behaviour. It is not an accident but dictated by the shape of the QCD-induced axion potential, which is ultimately a consequence of the fact that the axion breaks a \emph{compact} global symmetry group. It is interesting to note that there is a different, but related issue in the quartic axion term induced by this potential. The sign of this term is fixed as well and leads to a strictly attractive self-interaction with consequences for the axion dark matter halo model~\cite{Fan:2016rda}. 

\section*{Acknowledgement}
We thank Itai Bloch, Quentin Bonnefoy, Kai Bartnick, Sebastian Ellis, Gilad Perez, Inbar Savoray, Javi Serra, Konstantin Springmann, Stefan Stelzl and Andreas Weiler for useful comments and in particular for pointing out work on non-perturbative axion field configurations in the context of astrophysical observables. We acknowledge support by the UKRI future leaders fellowship DARKMAP.

\bibliography{biblio}

\end{document}